# First production and detection of cold antihydrogen atoms[*]

M.C. Fujiwara[a†], M. Amoretti[b], C. Amsler[c], G. Bonomi[d], A. Bouchta[d], P. Bowe[e], C. Carraro[bf] C.L. Cesar[g], M. Charlton[h], M. Doser[d], V. Filippini[i], A. Fontana[ij], R. Funakoshi[a], P. Genova[i], J.S. Hangst[e], R.S. Hayano[a], L.V. Jørgensen[h], V. Lagomarsino[bf], R. Landua[d], D. Lindelöf[c], E. Lodi Rizzini[k], M. Marchesotti[i], M. Macri[b], N. Madsen[e], P. Montagna[ij], H. Pruys[c], C. Regenfus[c], P. Rielder[d], A. Rotondi[ij], G. Testera[b], A. Variola[b], D.P. van der Werf[h]

(ATHENA Collaboration)

[a] Department of Physics, University of Tokyo, Tokyo 113-0033 Japan
[b] Istituto Nazionale di Fisica Nucleare, Sezione di Genova, 16146 Genova, Italy
[c] Physik-Institut, Zürich University, CH-8057 Zürich, Switzerland
[d] EP Division, CERN, Geneva 23 Switzerland
[e] Department of Physics and Astronomy, University of Aarhus, DK-8000 Aarhus, Denmark
[f] Dipartimento di Fisica, Università di Genova, 16146 Genova, Italy
[g] Instituto de Fisica, Universidade Federal do Rio de Janeiro, Rio de Janeiro 21945-970, Brazil
[h] Department of Physics, University of Wales Swansea, Swansea SA2 8PP, UK
[i] Istituto Nazionale di Fisica Nucleare, Sezione di Pavia, 27100 Pavia, Italy
[j] Dipartimento di Fisica Nucleare e Teorica, Università di Pavia, 27100 Pavia, Italy
[k] Dipartimento di Chimica e Fisica per l'Ingegneria e per i Materiali, Universit`a di Brescia, 25123 Brescia, Italy

**Abstract**

The ATHENA experiment recently produced the first atoms of cold antihydrogen. This paper gives a brief review of how this was achieved.

**1. Introduction**

In 1992, Dan Kleppner wrote in his summary of the Munich Antihydrogen workshop[1]:

> *"In the past 6 years, the creation of antihydrogen has advanced from the totally visionary to the merely very difficult."*

Although production of relativistic antihydrogen was reported a few years later[2,3], it took 10 years from Kleppner's remark before the first cold atoms of antihydrogen were produced and detected by the ATHENA collaboration[4]. Our results were subsequently corroborated by the ATRAP collaboration[5].

---

[*] Invited talk at International Conference on Low Energy Antiprotons.
[†] Corresponding author: e-mail: Makoto.Fujiwara@cern.ch

Two reports of high-energy antihydrogen production, first at CERN[2] in 1996 and later at FERMILAB[3] in 1998, showed the existence of neutral anti-atoms. However, these fast antihydrogen atoms were not suited for precision studies of neutral anti-matter, as they annihilated almost immediately after creation. The production of cold antihydrogen was, thus, eagerly awaited.

After the shutdown of the highly successful LEAR machine, in order to concentrate resources for the LHC construction, the AD facility was constructed at CERN with external funding, largely from Japan. The facility, after a few days of test beam in 1999, was commissioned for physics in July 2000. Two experiments, ATHENA and ATRAP, aim for the precision spectroscopy of antihydrogen, while another experiment (ASACUSA) has been studying spectroscopy of antiprotonic atoms and antiproton collisions.

By the end of 2000, the year in which the last LEAP conference was held, ATRAP had demonstrated positron cooling of antiprotons[6], while ATHENA had just learned to cool antiprotons with electrons[7]. Nonetheless, rapid progress was made, and in the end ATHENA's distinct features (Section 3) allowed us to produce and detect the first cold atoms of antihydrogen. In this article, we give a brief review of how we got there. First, though, we discuss some of the reasons for pursuing this goal.

## 2. Motivations for cold antihydrogen

Testing fundamental symmetries is an important subject in physics. Invariance of physical laws under the combined operations, taken in any order, of Charge-conjugation, Parity, and Time reversal (CPT), is guaranteed in local quantum field theories of point-like particles in flat space time by the CPT theorem[8] under assumptions including Lorentz invariance and unitarity. These assumptions, however, are not implicit in some classes of theories beyond the Standard Model. Recently, there is growing interest in CPT and Lorentz violation, and this is due in part to the development by Kostelecky and co-workers of an extension of the Standard Model[9] model that incorporates these violations. Ellis, Mavromatos, and co-workers have also proposed other scenarios of CPT violation involving Quantum Gravity[10].

Another notable recent development is the idea[11] of extra dimensions that are large compared to the Plank Scale ($10^{-33}$ cm). It was originally proposed, as an alternative to the popular super-symmetric theories, to solve the gauge hierarchy problem of the Standard Model. We could consider variation of these models which introduce a new gauge interaction in the "bulk", or compactified extra dimensions. Although detailed calculations need to be carried out, there may be measurable CPT violations at low energies, if extra dimensions, at the electroweak scale for example.

Cosmological baryon asymmetry in the Universe is normally associated with the famous Sakharov conditions involving CP violation (in baryons or leptons) and thermal non-equilibrium, but alternative models of baryogenesis are possible with CPT violation in thermal equilibrium[12,13,14]. CPT violation in the neutrino sector[15] is also an active area of current research.

There exist numerous experimental tests of CPT invariance[16], of which the most often quoted is that of the neutral kaon relative mass difference at the level of $10^{-18}$. Note however that Kobayashi and Sanda [17], as well as Bigi[18], have questioned the significance of dividing the possible mass difference with the mass itself. Given the fundamental importance of CPT symmetry, it should be tested in all particle sectors where precision results can be expected. Comparison of antihydrogen properties with those of its well-studied matter counterpart could provide a competitive direct test of CPT symmetry, despite the considerable technical challenges.

Gravitational acceleration on antiparticles has never been directly measured, though various indirect limits exist, for example, from weak equivalence principle tests[19]. Extremely cold antihydrogen, if produced, could be used for such tests.

## 3. ATHENA features

A distinctive feature of the ATHENA antihydrogen apparatus (Fig. 1) is its open and modular philosophy. By avoiding a sealed vacuum, a powerful method of positron accumulation using buffer gas[20,21] could be adopted while maintaining an excellent vacuum in which antiprotons survive for many hours[7]. The open system should allow relatively easy introduction of laser light, and provide a possibility of extracting antihydrogen atoms as a beam in the future.

A high-granularity position sensitive detector[22] was the key for unambiguous identification of the production of antihydrogen atoms by detecting the annihilations of antiprotons and positrons occurring at the same place and same time. Newly developed diagnostic techniques[23] allowed us on-line monitoring of the conditions of the trapped positron plasma during its interaction with the antiprotons, providing an ability to rule out alternative (yet unlikely) interpretations of data, given e.g. in Ref [5].

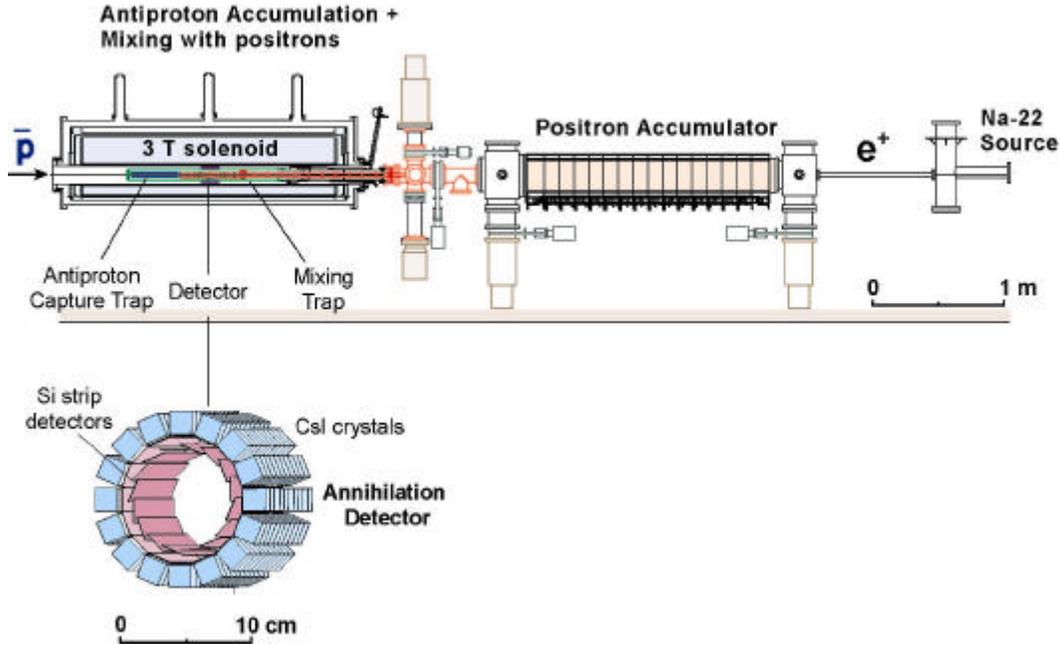

**Figure 1**. An overview of the ATHENA apparatus

## 4. Making the first cold antihydrogen

Antiprotons are obtained from the CERN's Antiproton Decelerator (AD) facility, which provides several times $10^7$ antiprotons about every 100 sec. After being slowed down in the degrader foils, antiprotons are dynamically trapped in a Penning trap, where their radial motion is confined by a 3T superconducting solenoid magnetic field, and the axial motion by a -5 kV potential well. These trapped antiprotons are cooled by their interaction with preloaded electrons, which in turn self-cool via the emission of synchrotron radiation. Figure 2 summarizes the result of the process of electron cooling of antiprotons.

Antiprotons from three AD spills are accumulated, for each mixing cycle, in the catching trap before they are transferred to the adjacent mixing trap. The process of antiproton transfer leaves room for improvement for better efficiency, and we will work on this in the near future. At the moment, about $10^4$ antiprotons (per cycle) are used for mixing with positrons.

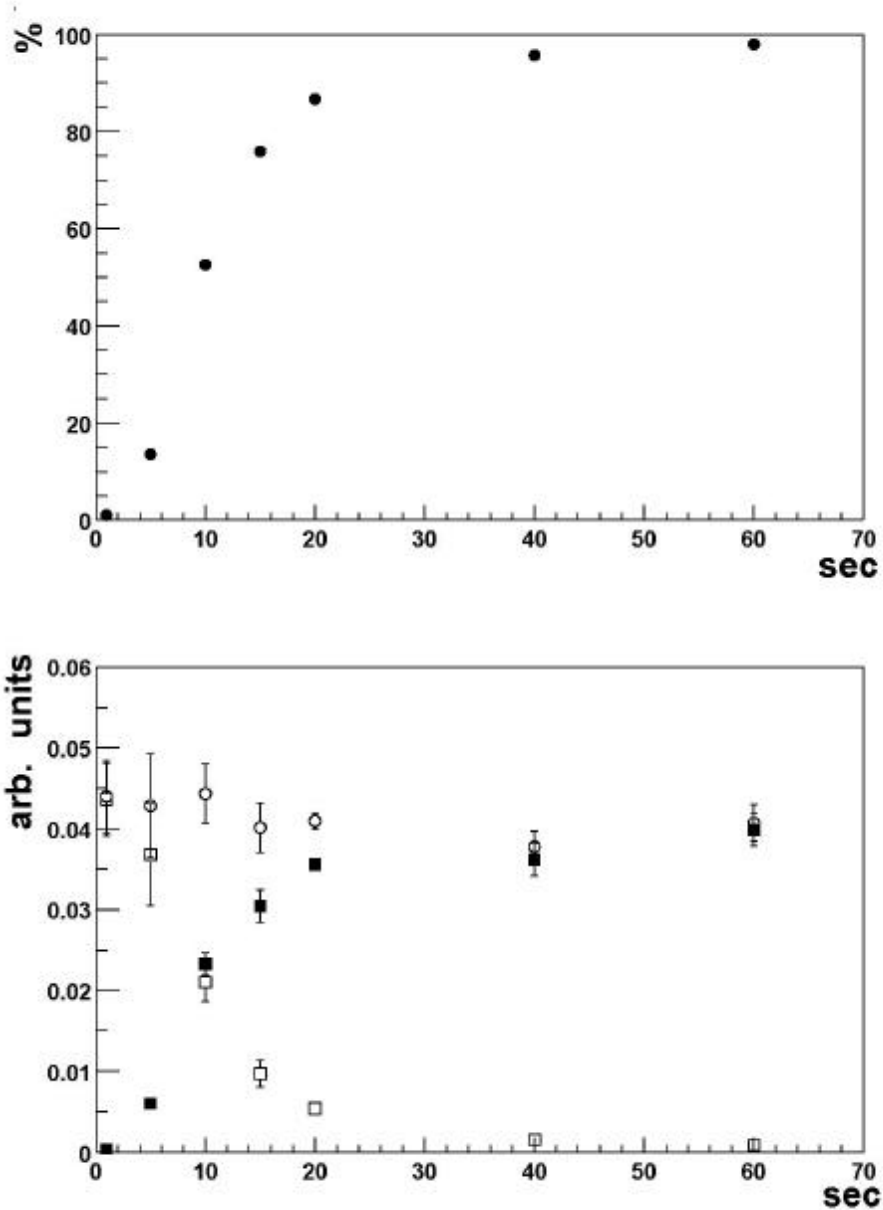

**Figure 2**. **Top**: The ratio of cooled (<1 eV) antiprotons to the sum of cooled and uncooled (~keV) ones is plotted as a function of cooling time, illustrating electron cooling efficiency. **Bottom**: The number of cooled (filled square), uncooled (open square) and cooled+uncooled (open circle) antiprotons are plotted as a function of cooling time. The data are normalized to the beam intensity, measured by scintillators read out by hybrid photodiodes[24]. For both Top and Bottom figures, the numbers of cooled and uncooled antiprotons were measured by extracting the antiprotons by slowly ramping down the potential wall, and counting their annihilation on the wall with scintillators. (Figure from Ref. [7].)

Positrons are obtained from a 40 mCi $^{22}$Na source, and are moderated by a frozen neon film. Their trapping and accumulation are achieved with the help of nitrogen buffer gas, which provides the dissipative process necessary for trapping the continuous flow of positrons. This is a method pioneered by Surko and coworkers[20]. About 150 million positrons are accumulated in a 0.14 T field every 5 min, they are then injected and re-trapped in the 3T field with a 50% efficiency. The accumulation rate of cold positrons in our mixing trap, normalized to the source strength, is $2.3 \times 10^7$ e$^+$ h$^{-1}$ mCi$^{-1}$, more than three orders of magnitude larger than the maximum rate reported using an alternative scheme[6].

The formation of antihydrogen takes place in an inverted double trap where the positron trap is housed inside the antiproton trap, a configuration referred to as a nested trap[25]. Mixing of antiprotons and positrons is initiated by injecting $10^4$ antiprotons into the cloud of $0.7 \times 10^8$ positrons trapped in the central interaction region. The injected antiprotons rapidly lose their energy via Coulomb collisions with positrons, as illustrated in Fig. 3. Here, the energy of antiprotons in the mixing trap is analyzed by gradually reducing the confining potential well, and detecting the annihilations of axially leaked antiprotons. Positron cooling of the antiprotons was first reported by ATRAP[6] with 300 times fewer positrons than we report here. Taking advantage of our powerful positron accumulator, we were able to perform systematic measurements under very different conditions with a large number of positrons[26].

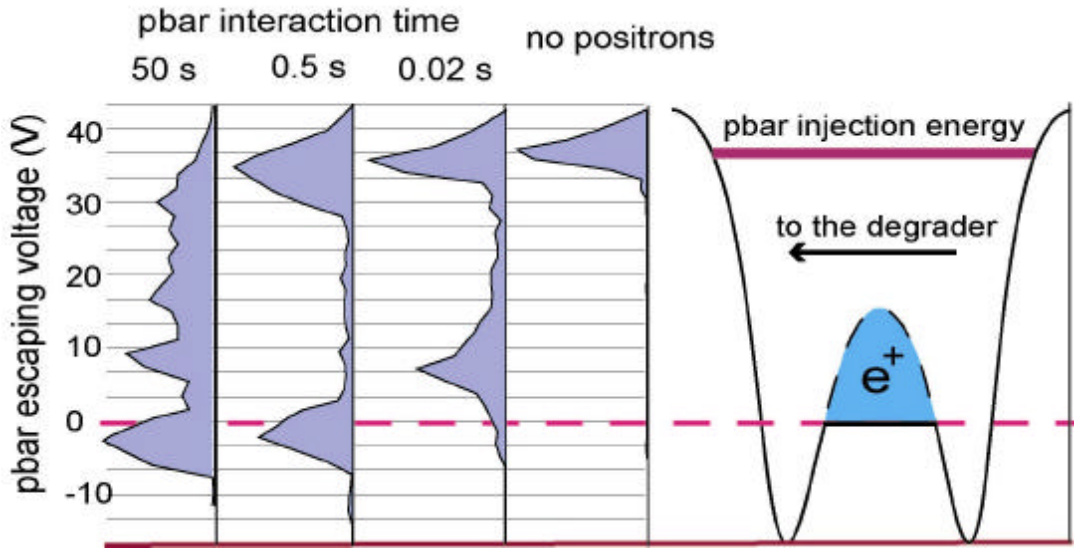

**Figure 3**. Cooling of antiprotons through contact with positrons in the mixing trap. Antiprotons are injected into the positron plasma at t=0. After a defined interaction time t (=0.02, 0.5, 50 s), the left potential well is ramped down, allowing the antiprotons to escape towards the degrader foil. The time distribution of the subsequent annihilations is converted to the distribution of the antiproton escaping voltage, using the known time evolution of the potential well. This indicates the antiproton energy distribution in the mixing trap after the given interaction time.

## 5. Antihydrogen signal

Neutral antihydrogen atoms, when formed, escape the electromagnetic confinement of the Penning trap, and drift until they reach the trap electrodes. Antihydrogen atoms were identified by detecting the annihilations of antiprotons and positrons, coincident in both position and time, and taking place at the electrode wall. Antiprotons annihilate into several charged or neutral particles (mostly pions), and the annihilation vertices were reconstructed with a precision of ~4 mm (1$\sigma$) by tracking the charged trajectories with two layers of double-sided silicon microstrip detectors. Back-to-back 511 keV gamma rays from positron annihilations were detected by highly segmented pure CsI crystals, read out by avalanche photodiodes. In order to suppress the background, we demand in the analysis that the crystal hits are isolated (i.e. no hits in the eight neighboring crystals), and have no charged track passing through. Gamma energy window was set around 511 ± 165 keV, based on channel-by-channel *in situ* calibration using positron annihilation in the trap.

We plot in Fig. 4 (a) the angle $\theta_{\gamma\gamma}$ between two gammas, seen from the reconstructed charged particle vertex. For the real antihydrogen events, there should be a peak at $\cos(\theta_{\gamma\gamma})\sim-1$, and indeed this is what we observed (histogram).

The background was carefully studied in several ways: (1) measurement with a heated positron plasma in which antihydrogen production is suppressed [Fig. 4a: triangle], (2) measurement without positrons and only antiprotons annihilating on the electrode wall [Fig. 4b: histogram], (3) standard mixing data analyzed with the gamma energy cut displaced above 1 MeV [Fig. 4b: circle]. In all the background cases, no peak at $\cos(\theta_{\gamma\gamma})\sim-1$ is observed, as expected. Note that the three-dimensional imaging capability of the antiproton annihilation, as well as high angular resolution for gamma detection with segmented crystals, were essential in discriminating against the angular-uncorrelated gamma background, which comes predominantly from the decay and the subsequent electromagnetic shower of neutral pions. These can produce secondary positrons (hence real 511 keV gammas) as well as higher energy gammas.

After our report was published, some doubts on our results were expressed[5,27]. For example, Ref. [27] argues that our background measurements are invalid and suggests an improved control measurement with antiprotons only. But this is indeed what we had done in (2) above, as we reported in Ref. [4].

The possibility of ambipolar diffusion[28] as an alternative interpretation of our results[5,27], is very unlikely due to the large ratio of the numbers of positrons to antiprotons under our condition (although this is not necessarily the case for the condition of Ref. [5]) and had been excluded by our on-line monitoring of the positron plasma evolution during the mixing process.

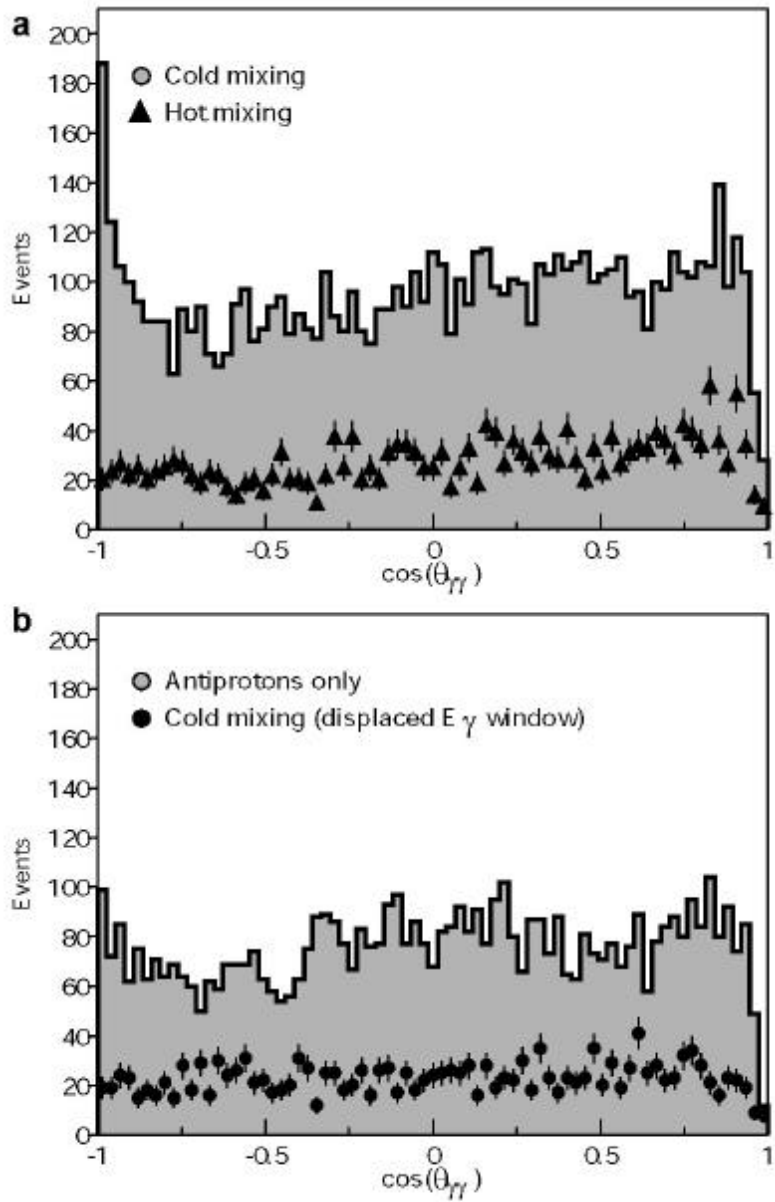

**Figure 4:** The angle between two 511 keV gamma rays, as seen from the reconstructed vertices of antiproton annihilations **(a):** antiproton mixing with cold positrons (grey histogram), and that with positrons heated with RF to several 1000 K (triangle). **(b):** without positrons, and only antiprotons annihilating on the electrode wall (grey histogram), and standard cold mixing data but analyzed with a gamma energy window displaced to a region above 511 keV (circle). (Figure from Ref. [4].)

## 6. More cold antihydrogen

In our first report[4], we adopted a very conservative approach, and focused only on the fully reconstructed events, i.e. the peak in the opening angle plot with $\cos(\theta_{\gamma\gamma})<-0.95$, and assumed that all of the flat part of the $\cos(\theta_{\gamma\gamma})$ plot is background. Based on these "golden" 131±22events, we gave a lower limit of 50000 antihydrogen produced, taking into account a conservatively estimated detection efficiency.

Our subsequent analysis of the charged vertices, combined with detailed Monte Carlo studies, are consistent with a substantial fraction of charged triggers being due to antihydrogen. Preliminary results suggest that in 2002 ATHENA produced of the order of one million antihydrogen atoms, a number substantially larger than produced in any other anti-atom experiment, or those combined.

## 7. Summary

In this article, we have reviewed the first production and detection of cold antihydrogen by the ATHENA collaboration. Beyond the first production, the versatile, high-duty cycle ATHENA apparatus has allowed a variety of studies using cold antihydrogen. We have only touched on some here. Others include the temperature dependence of formation, the temporal modulation of antihydrogen production, and antihydrogen emission angular distribution, and are the subject of forthcoming articles.

The ultimate goal of precision spectroscopy of antihydrogen is as yet *very difficult*, but hopefully not *totally visionary*. Physics with cold antihydrogen has just begun, and we have much to look forward to.

## Acknowledgements

We thank J. Rochet, S. Bricola, H. Higaki, Y. Yamazaki, G. Sabrero, and P. Chiggiato for their valuable contributions, and CERN's PS and AD crew for providing the excellent antiproton beam. We gratefully acknowledge the financial support from INFN (Italy), FAPERJ (Brasil), MEXT (Japan), SNF (Switzerland), NSRC (Demark), EPSRC (UK), and European Union.